\vsize=23.5truecm \hsize=16truecm
\baselineskip=0.5truecm \parindent=0truecm
\parskip=0.2cm \hfuzz=1truecm

\font\scap=cmcsc10

\font\tenmsb=msbm10
\font\sevenmsb=msbm7
\font\fivemsb=msbm5
\newfam\msbfam
\textfont\msbfam=\tenmsb
\scriptfont\msbfam=\sevenmsb
\scriptscriptfont\msbfam=\fivemsb
\def\Bbb#1{{\fam\msbfam\relax#1}}

\newcount\eqnumber
\eqnumber=1
\def\neweq{{\rm{(\the\eqnumber)}}\global\advance\eqnumber by 1}
\def\eqdef#1{\eqno\xdef#1{\the\eqnumber}\neweq}
\def\newaeq{{\rm{(\the\eqnumber a)}}\global\advance\eqnumber by 1}
\def\eqdaf#1{\eqno\xdef#1{\the\eqnumber}\newaeq}
\def\eqdisp#1{\xdef#1{\the\eqnumber}\neweq}
\def\eqdasp#1{\xdef#1{\the\eqnumber}\newaeq}

\newcount\refnumber
\refnumber=1
\def\newref{{\the\refnumber}\global\advance\refnumber by 1}
\def\refdef#1{{\xdef#1{\the\refnumber}}\newref}

\newcount\fignumber
\fignumber=1
\def\newfig{{\the\fignumber}\global\advance\fignumber by 1}
\def\figdef#1{{\xdef#1{\the\fignumber}}\newfig}

\def\smallskip{\vskip 3pt}
\def\medskip{\vskip 6pt}
\def\bigskip{\vskip 12pt}

\def\bol{\scalebox{0.7}{$\bullet$}}

\input graphicx

\centerline{\bf The trouble with deautonomising higher order maps}
\bigskip
\medskip{\scap R. Willox}  \quad
{\sl Graduate School of Mathematical Sciences, the University of Tokyo, 3-8-1 Komaba, Meguro-ku, 153-8914 Tokyo, Japan}
\medskip{\scap B. Grammaticos} and {\scap A. Ramani} 
\quad{\sl Universit\'e Paris-Saclay and Universit\'e de Paris-Cit\'e, CNRS/IN2P3, IJCLab, 91405 Orsay, France}

\bigskip
{\sl Abstract}
\smallskip
The deautonomisation of birational maps that have the singularity confinement property, i.e. the construction of nonautonomous versions of such maps that preserve the singularity properties of the original, has proven crucial in our understanding of the mathematical properties  behind the integrability of second order maps. For example, the deautonomisation procedure led directly to the development of a general theory of discrete Painlev\'e equations, and it seems highly likely it will play a crucial role in any future theory of higher dimensional Painlev\'e equations as well. Generally speaking however,  higher order integrable mappings may have non-confined singularities and it is important to understand if, and how, deautonomisation should work for such mappings. In this paper we explore different deautonomisation scenarios on a series of carefully constructed higher order mappings, integrable as well as non-integrable, that possess non-confined singularities and we challenge some common assumptions regarding the co-dimensionality of the singular loci that might play a role in the deautonomisation process. Along the way we also propose a novel procedure to calculate the growth of the multiplicities of singularities that appear in so-called anticonfined singularity patterns, based on an ultradiscrete version of the mapping.

\bigskip
Mathematics Subject Classification: 37J70, 14E05
\smallskip
Keywords: integrability, integrable maps, deautonomisation, singularities, discrete Painlev\'e equations
\bigskip

\bigskip
1. {\scap Introduction}
\medskip
The deautonomisation procedure, based on singularity confinement, was introduced (although not yet referred to by that name) immediately upon the proposal of singularity confinement as an integrability criterion [\refdef\singconf]. The procedure consists in obtaining integrable {\sl non-autonomous} extensions of integrable {\sl autonomous} mappings such that  these extensions preserve the singularity properties of the original maps. The main application of the deautonomisation procedure was the derivation of the discrete analogues of the Painlev\'e equations, the vast majority of which were obtained by the deautonomisation of mappings belonging to the Quispel-Roberts-Thompson (QRT) family [\refdef\qrt]. More recently, deautonomisation based on singularity confinement was also shown to offer a reliable and rigorous way to obtain the dynamical degree (or the algebraic entropy) of any three point mapping that has confined singularities [\refdef\redemption,\refdef\alex].

The deautonomisation procedure for a birational map $f$ (on $\Bbb P^1 \times \Bbb P^1$ or $\Bbb P^2$), intrinsically, relies on the fact that saying that a map has the singularity confinement property is equivalent to saying that it can be regularized to an automorphism on a suitable rational surface, by a finite number of blow-ups [\refdef\dilfav]. Such a rational surface is often referred to as a space of initial conditions for the original map. Deautonomizing a map that has a space of initial conditions---i.e. that has the singularity confinement property---then works by requiring that the nonautonomous map should also posses a space of initial conditions, which now  takes the form of a family of isomorphic rational surfaces, one for each iterate of the map [\refdef\maseth]. As was shown in [\refdef\royal], if the deautonomisation preserves all the singularity patterns of the original autonomous map, then certain dynamical properties such as for example its dynamical degree $\lambda_*$, 
$$\lambda_* = \lim_{n\to\infty} (\deg f^{(n)})^{1/n},\eqdef\dyndeg$$
which captures the growth rate of successive iterates $f^{(n)}$ of the map  (or, equivalently, its algebraic entropy $\varepsilon=\log\lambda_*$), will be identical to those of the original autonomous map one started from. Note that in many cases however, it is possible to postpone the exact instance (i.e. iteration) at which a particular singularity for the nonautonomous map will confine, while still keeping the singularity confinement property for the map as a whole. This process is referred to as `late deautonomisation'---first proposed in [\refdef\hivtwo], and its algebro-geometric foundation first explained in [\royal]---and in this case the dynamical properties of the resulting non-autonomous map do differ from those of the original autonomous one.

The dynamical properties of autonomous birational maps $f$ on $\Bbb P^1 \times \Bbb P^1$ that possess a space of initial conditions were classified in [\dilfav]. Such a map has either:\hfill\break
{\bf (i)} bounded degree growth for its iterates (and hence $\lambda_*=1$), in which case it is periodic or (as shown in [\refdef\blancdeserti]) it is birationally equivalent to a projective map on $\Bbb P^2$\hfill\break
{\bf (ii)} quadratic degree growth for its iterates (and hence $\lambda_*=1$), in which case it necessarily preserves an elliptic fibration\hfill\break
{\bf (iii)} exponential degree growth (the degree of its iterates grows asymptotically as $\lambda_*^n$ for $\lambda_*>1$), in which case it does not preserve any fibration at all and (as shown in [\refdef\cantat])  any birational map $g$ that would commute with it, i.e. $g\circ f = f\circ g$, is necessarily trivially related to it:  $^\exists n\in{\Bbb Z}, m\in{\Bbb Z}_{>0}: g^{(m)}=f^{(n)}$.\noindent\par
In [\maseth] similar statements pertaining to degree growth were proven for nonautonomous maps on  $\Bbb P^1 \times \Bbb P^1$ that possess a space of initial conditions (but not the remaining statements). In the same paper it is also shown that any nonautonomous mapping with a space of initial conditions that is obtained from a late deautonomisation, necessarily exhibits asymptotic exponential growth for its iterates.

In case an autonomous birational map  on $\Bbb P^1 \times \Bbb P^1$ does {\sl not} possess a space of initial condtions, it is shown in [\dilfav] that it must either have the above property {\bf (iii)}, or that it must exhibit linear degree growth for its iterates, in which case it necessarily preserves a rational fibration and, typically, will be integrable through some direct linearisation procedure.

Which brings us to the general notion of {\sl integrability} for birational maps (on $\Bbb P^1 \times \Bbb P^1$ or $\Bbb P^2$). From the classification results presented above it is clear that besides the trivial case {\bf (i)},  the two 
 cases that deserve to be called `integrable' in the sense that they have some nontrivial invariance property that would allow one to integrate them, both exhibit non-exponential degree growth for their iterates: quadratic degree growth if the mapping has a space of initial conditions, or linear degree growth if it does not. Hence it seems natural, at least for autonomous mappings, to characterize (or, if one chooses to do so, to `define') integrable, autonomous, second order birational maps as maps that have non-exponential growth and, hence, a dynamical degree that is equal to 1 (and  therefore zero algebraic entropy). This `slow growth' definition of integrability, which goes back to the algebraic complexity arguments of Arnol'd [\refdef\arnold] and Veselov [\refdef\veselov], has now become a widely adopted definition of integrability,  even for nonautonomous or higher order maps for which no general results on the existence of invariants etc. are known. In the following we shall also adopt this definition and call any birational map that has zero algebraic entropy integrable. Note that this  also includes periodic maps (or any other map with bounded growth). In the following, however, we shall only consider infinite order maps and leave periodic maps aside.

It should also be noted that  linearisable maps (those with linear degree growth in the above classification) can, by their very nature, include arbitrary functions that depend on the exact iterate at which the map is evaluated---which could be viewed as a deautonomisation of sorts---but that the deautonomisation procedure based on singularity analysis, described above, has only ever been defined for maps that possess a space of initial conditions, i.e. whose singularities enjoy the singularity confinement property. Which is not the case for linearizable maps on $\Bbb P^1 \times \Bbb P^1$ (or $\Bbb P^2$).

The general adoption of sub-exponential growth as a practical definition of integrability for (bi-)rational maps, in any dimension, has of course intensified the search for simple yet rigorous methods for calculating the dynamical degree (or algebraic entropy) of any given map. For mappings on $\Bbb P^1 \times \Bbb P^1$ (or $\Bbb P^2$) a rigorous procedure for calculating the dynamical degree by successive blow-ups was introduced by Takenawa in [\refdef\takenawa] for mappings that possess a space of initial conditions and, in general, by Diller and Favre in [\dilfav]. In [\refdef\bedkim], Bedford and Kim describe a similar procedure for birational mappings on $\Bbb P^N$ when $N>2$, in case the mapping at hand can be regularized to a so-called pseudo-isomorphism of surfaces (i.e. an isomorphism in co-dimension 1). As far as we know, no generally applicable, blow-up based, computational procedure is known for general birational mappings on  $\Bbb P^N$, for general $N$. In fact, the recent discovery of a birational map on $\Bbb P^3$ with a transcendental dynamical degree [\refdef\belletal] seems to suggest that a general procedure that would be guaranteed to produce the exact value of the dynamical degree for any given (bi-)rational map most probably does not exist. Nonetheless, several non-blow-up based computational approaches are known that yield an explicit minimal polynomial for the dynamical degree of birational maps on $\Bbb P^N$, their singularity structure allowing. See for example [\refdef\alonsoo,\refdef\alonsot] or [\refdef\vialleto, \refdef\viallett].

The situation is far simpler for birational mappings on $\Bbb P^2$ (or $\Bbb P^1 \times \Bbb P^1$), for which the dynamical degree is necessarily an algebraic integer [\dilfav], and for which there exist non-blow-up based procedures that have been shown to yield the exact value of the dynamical degree (through its minimal polynomial, which is calculated explicitly). One such method is due to Halburd [\refdef\hardrod], in which successive iterates of the map are treated as functions of a single variable. In that  case, the elementary fact that the degree of an iterate of the map is actually given by the number of preimages of the rational function (in the chosen variable) obtained at that iterate, for any value that function can take on $\Bbb P^1$, combined with the idea that the numbers of preimages for values that appear in the singularity patterns of the map will be easily related to each other and will lead to linear recurrence relations for the degrees, yields a highly efficient method for obtaining minimal polynomials for dynamical degrees. Unfortunately it is not guaranteed in general that the singularity patterns yield enough information to construct such recurrences (see e.g. [\refdef\express] for a counterexample), but in those cases where the singularity patterns do form a closed system of equations, the method has been shown to yield the exact value of the dynamical degree [\refdef\maserod]. The method can however also be applied to mappings with non-confined singularities (see e.g. [\refdef\nonconfrod]) and has been shown to work for certain types of higher order maps as well [\refdef\higher], although a firm mathematical justification for these surprising facts is still lacking. In the following sections we will see several examples of higher order mappings for which Halburd's method does seem to work.

Another remarkably simple method, first  described in [\redemption], that has been shown to yield the exact value of the dynamical degree, at least for three-point mappings that possess a space of initial conditions, relies on the deautonomisation of (a possibly extended version) of a given (autonomous) map. In [\alex] it has been shown that, under certain conditions on the map, the degree growth of its iterates is reflected in the asymptotic behaviour of the coefficient functions in a sufficiently rich deautonomisation of that map. However, although the proof given in [\alex] relies heaviliy on the fact the map has the singularity confinement property, and only works for second order maps, the method itself still seems to work for some second order mappings that have non-confined singularities besides confined ones and, especially, also for higher order maps (see e.g. [\higher]). The fact that the full-deautonomisation method seems to be applicable to higher order maps as well, is one of the main motivations to study the problem of deautonomizing higher order maps in more generality.

However, as we have shown in [\refdef\premierpapier], at higher orders, even integrable maps can have non-confined singularities, e.g. when they exhibit polynomial degree growth that is faster than quadratic. Hence, if one believes that a method like the full-deautonomisation method might provide a useful tool even for such higher order maps, one is forced to tackle the problem of how to deautonomise maps that have only a few confined singularities, besides a plethora of non-confined ones, or even none at all. This is the main motivation for the work presented in this paper.
As explained  in [\premierpapier], it is possible to construct integrable mappings with unconfined singularities by coupling lower order integrable mappings with confined singularities, to linearisable ones. In this paper, using some carefully crafted examples of such coupled maps, we shall present several different scenarios in which information obtained from the singularity patterns (be they confined or not) can be used to successfully deautonomise the mapping. Moreover, as we shall see, a successfull deautonomisation is not only required to preserve the singularity structures or patterns of the original autonomous mapping, but should also yield a mapping that has the same degree growth as the original one, a fact which at higher order is no longer guaranteed by simply preserving a singularity pattern. 

For this reason we shall have to ascertain the precise degree growth of the (higher order) mappings we consider, and of their deautonomisations. To this end, we shall essentially use two methods. Halburd's method, which was explained above (or a simplified version thereof  [\express], which only yields the value of the dynamical degree of the map and not the exact degree sequence), or for cases with exponential degree growth---i.e.non-integrable cases---the so-called Diophantine method [\refdef\diophantine] which provides a simple and (usually) computationally efficient way to approximate the numerical value of the dynamical degree. We shall use the Diophantine method as a test, to check whether the numerical value it yields matches that obtained from Halburd's method (or its simplified form), which would suggest that it coincide with the value of the dynamical degree of the mapping.

However, it is important to point out that the Diophantine method actually does not approximate the dynamical degree directly, but rather a different characteristic of the dynamics of the map, its so-called {\sl arithmetic degree} (see e.g. [\refdef\silvo]):
$$\lambda_\alpha = \lim_{n\to\infty} h_{\Bbb Q}(f^{(n)})^{1/n},\eqdef\algdeg$$
where $h_{\Bbb Q}: {\Bbb Q}\to [1,\infty[$ is some well-defined height function over the rationals, for example the maximal number of digits in the numerator and denominator of the iterates $f^{(n)}$ of $f$, calculated as a reduced fraction for rational initial conditions (and, practically speaking, for appropriately general rational values of any parameters that might appear in the mapping). Note that, contrary to the limit (\dyndeg) in the definition of the dynamical degree, the existence of the limit in the definition (\algdeg) is not an established fact (no counterexamples are known to this date, but no (general) proof of its existence is known either). In [\refdef\kawasilv] it is shown that if the limit in (\algdeg) exists, then necessarily
$$\lambda_\alpha\leq\lambda_*,$$
and it is conjectured (in the same paper) that in that case the arithmetic and dynamical degrees actually coincide, but to this date this has been proven only in special circumstances (see e.g. [\refdef\silvt] or [\refdef\matsu]). Nonetheless, in the following, for nonintegrable mappings, we shall calculate their arithmetic degree to see if it matches our own prediction of the dynamical degree (through Halburd's method), assuming naively  that coincidence of two different conjectures in some sense corroborates both.

In the following sections we shall construct several examples of higher order maps, by coupling known  integrable or non-integrable second order mappings with linear ones, such as to obtain mappings that exhibit specific types of singularity patterns that sometimes necessitate slightly different and at times very different deautonomisation strategies. Finally, in the conclusions we shall point out the apparent limits to this attempt at deautonomising nonconfining higher order mappings, which we believe is a required first step in the exploration of the properties of nonautonomous versions of mathematically and physically interesting higher order maps.

\bigskip
2. {\scap Deautonomising higher order  integrable mappings}
\medskip
We start our exploration of possible deautonomisation procedures for higher order mappings with the case of integrable mappings. As mentioned at the end of the previous section, we wish to describe different scenarios that can arise when trying to deautonomise a given higher order mapping and we shall construct examples tailored especially to each individual scenario. The examples in this section will all be obtained by coupling  the (integrable) QRT mapping
$$x_{n+1}x_{n-1}=1-{a\over x_n},\eqdef\bzena$$
with different types of linear, or linearisable, mappings, resulting in a higher order mapping with the desired properties. Note that, by construction, the obtained mapping should be considered to be integrable. In the following we shall, implicitly, think of  (\bzena) as  a map $(x_{n-1}, x_n)\to (x_n,(x_n-a)/(x_x x_{n-1}))$ on $\Bbb P^1\times \Bbb P^1$, but we shall keep on using the equation notation since the results of the singularity confinement calculations are easier to present that way.

Here $a$ is an arbitrary non-zero (constant) complex number and the mapping has two singularities, i.e. iterates at which the inverse mapping is not defined: at $x_n=a$ and $x_n=0$, for generic values of $x_{n-1}$.
We shall not go into all the detail of the calculations involved in checking the singularity confinement property (a detailed explanation of the method can be found e.g. in [\refdef\eulerbook]) and we only summarize the results. For the first singularity, if one starts from $x_{n-1}=u, x_n=a$, at some $n$ for some generic value of $u\in{\Bbb C}$, we find that $x_{n+1} = 0, x_{n+2}=\infty, x_{n+3}=\infty, x_{n+4}=0$, at which point the mapping becomes indefinite. However, it so happens that if one performs the same calculation for $x_n=a+\epsilon$, for some non-zero $\epsilon$ and then takes the limit $\epsilon\to0$ for each iterate, one finds that $x_{n+5}=a$ and $x_{n+6}=u$, after which subsequent iterates will no longer hit any singularities (assuming that $u$ is indeed generic). We say that this singularity is {\sl confined} and its behaviour can be summarized in the singularity pattern

$$\{a,0,\infty,\infty,0,a\}.\eqdef\spone$$

The second singularity is of a different type. Starting from $x_{n-1}=u, x_n=0$, again for generic $u\in{\Bbb C}$, one finds $x_{n+1}=\infty, x_{n+2}=\infty, x_{n+3}=0$ and one hits the same indefiniteness as before. However, performing a similar calculation as above one finds that $x_{n+4}=a^2/u, x_{n+5}=\infty, x_{n+6}=u/a^2$ and $x_{n+7}=0$ and this whole sequence of iterates is repeated again, mutatis mutandis at $x_{n+11}, x_{n+13},  {\rm  etc...}$. Given the form of equation (\bzena) it is clear that for our initial conditions, the inverse map necessarily yields $x_{n-1}=\infty$, and that the pattern we found therefore also repeats indefinitely (but in reverse order) towards decreasing values of $n$. This type of singularity is called {\sl cyclic} (in this case with period 7) and it is known to be completely inconsequential when it comes to integrability issues. Henceforth we shall neglect any cyclic patterns that might arise for the mappings that we construct.

In order to deautonomise (\bzena) based on the confined singularity pattern (\spone),  we assume that the parameter $a$ in equation (\bzena) is no longer constant, but a function of the independent variable: $a_n$. Starting the singularity confinement calculation with $x_n=a_n$ instead of $a$, we find that in order to have $x_{n+5}=a_{n+5}$ and preserve the singularity pattern (\spone) by avoiding the emergence of an infinity at the next iterate, $a_n$ must satisfy the relation 
$$a_{n+5}a_n=a_{n+4}a_{n+1}.\eqdef\bzdyo$$
The solution of this simple linearizable recursion relation  is $a_n=\lambda^n\psi_4(n)$, for some $\lambda\in{\Bbb C}\setminus\{0\}$ and where $\psi_m$ denotes a general periodic function with period $m$. After removing some spurious degrees of freedom by gauge transformation of $x_n$ we thus obtain the well-known $q$-Painlev\'e equation [\refdef\asymm],
$$x_{n+1}x_{n-1}=1-{\lambda^n\psi_2(n)\over x_n},\eqdef\qP$$
as a natural non-autonomous version of the QRT map (\bzena). By construction, both equations have the singularity confinement property with essentially the same singularity patterns (including the cyclic pattern). This implies that they both possess a space of initial conditions and exhibit the same quadratic growth for their iterates. The essential difference between them being that the original QRT map has an invariant and can be integrated in terms of elliptic functions, whereas the general solution of the $q$-Painlev\'e equation (\qP) has no rational invariant and is thought to be transcendental.

We shall now start constructing higher order maps by coupling the nonautonomous equation
$$x_{n+1}x_{n-1}=1-{a_n\over x_n},\eqdef\orgeq$$
with essentially three different types of linear or linearizable maps, initially without assuming anything about the coefficient function $a_n$ in (\orgeq),  and we shall try to see if and how one can obtain deautonomisations of the resulting higher order maps by imposing constraints on their singularity patterns. These should then give rise to constraints on the functions $a_n$, which should of course be compatible with relation (\bzdyo). Some of the mappings we shall encounter possess only non-confined singularities and we shall thus be confronted with the problem of defining deautonomisation based on non-confined singularity patterns, which is completely unchartered territory.

One last important point that needs to be stressed concerns the notion of `singularity' for higher order maps. In the spirit of [\bedkim] and in keeping with our earlier work on the degree growth of higher order mappings [\premierpapier,\higher,\refdef\fast], we will limit ourselves to singularity patterns that originate in a singular locus of co-dimension 1 and we will completely neglect all singular behaviour due to loci with greater co-dimension when studying the singularity properties of the maps that we shall construct. Possible side effects of this limitation will be addressed in the conclusions.

\medskip
2.1. {\sl Scenario I -- Standard deautonomisations based on confined patterns and variations thereof}
\smallskip
Let us first couple equation (\orgeq) to the linear mapping
$$y_{n+1}=x_ny_n,\eqdef\bztri$$
which leads to the third order non-autonomous equation
$$y_{n+1}={y_{n-2}\over y_{n-1}}\left(y_n-a_{n-1}y_{n-1}\right).\eqdef\bztes$$ 
In the autonomous case, i.e. when $a_n$ is constant, we start with (generic) initial conditions $y_0, y_1$ and $y_2=a y_1$ and obtain the singularity pattern
$$\{a y_1, 0,*,\infty,*,*\},\eqdef\sptwo$$
where $*$ indicates a finite non-zero value where some degrees of freedom are missing. In the case at hand we find that $y_6$ and $y_7$ are functions only of $y_1$ but that $y_0$ reappears  in $y_8$. Thus the singularity is confined, in the sense that the freedom lost at $y_3$ has been recovered, but only at the level of $y_8$. When $a_n$  is not constant  we enter the singularity through $y_2=a_1y_1$ and after a simple calculation we find that $y_8$ diverges unless $a_n$ satisfies the condition (\bzdyo). 
Hence, in this case the deautonomisation procedure unfolds exactly as in the standard second order case: the correct deautonomisation constraint is obtained solely from the confined singularity pattern (\sptwo), which only contains the values $0, \infty$ and some regular values that depend on the initial conditions $y_0$ and $y_1$.  The important observation here, however, is that the singularity pattern for the original autonomous case on which the deautonomisation is based, should be calculated carefully, bearing in mind that `confinement' only really occurs at the instant where the sequence of iterates again returns to a co-dimension 1 variety.

Both the nonautonomous map (\bztes) with $a_n$ subject to condition  (\bzdyo), as well as its original autonomous version exhibit the same quadratic degree growth. For example, starting from generic $y_0$ and $y_1$ and computing the degree of the iterates in terms of $y_2$ we find 0, 0, 1, 1, 1, 2, 3, 4, 5, 6, 8, 9, 11, 13, 15, 17, 19, 22, 24, 27, 30, 33, 36, 39, 43, 46, $\cdots$, which clearly grows quadratically.

That equation (\bztes) is the result of a coupling to a linear equation manifests itself in the appearance of a so-called anticonfined singularity pattern  [\refdef\anticonf]:
$$\{\cdots,0,0^2,0^2,0,0,0^2,0,\bol,0,0,\bol,\bol,0,\bol,\infty,\bol,\bol,\infty,\infty,\bol,\infty,\infty^2,\infty,\infty,\infty^2,\infty^2,\infty,\infty^2,\infty^3,\cdots\}.\eqdef\acpone$$
In the above pattern (and hereafter) $\bol$ represents a free regular value, which is neither $0$ nor $\infty$. Furthermore, the exponents in the pattern above have the following meaning: if we had introduced $\epsilon$ in lieu of 0 in the calculation of the above pattern, we would have obtained iterates proportional to $\epsilon^k$ and $\epsilon^{-k}$, which we denote here by $0^k$ and $\infty^k$, respectively; in the following we shall refer to the exponent $k$ in this notation as the {\sl multiplicity} for that entry in the pattern. 

Note that the above pattern contains consecutive $\bol$ at two different places, which means there are two different ways of choosing initial conditions so as to end up in this particular singularity pattern.
One can either enter the pattern through the initial condition $(y_0,y_1,0)$ in which case, iterating towards increasing $n$, one obtains the sequence $(y_0, y_1, 0,-a_1y_0,\infty,y_1,a_1a_4y_0,\infty,\cdots)$, or through $(y_0,y_1,\infty)$ in which case the preceding iterations are $(\cdots,y_1/(a_{-1}a_{-4}),y_0,0,-y_1/a_{-1},\infty,y_0,y_1,\infty)$. It goes without saying that since the growth of the multiplicities in the anticonfined pattern is only linear, the presence of this pattern does not have any influence on the fact that the degree growth of  (\bztes) is quadratic and, hence, has no bearing on its integrability. 

As it turns out, we can introduce a further linear coupling by 
$$z_{n+1}=y_nz_n,\eqdef\bzpen$$
without changing the situation too much.
The resulting equation has the form
$$z_{n+1}={z_nz_{n-2}\over z_{n-1}^2z_{n-3}}\left(z_nz_{n-2}-a_{n-2}z_{n-1}^2\right),\eqdef\bzhex$$
which we shall study without assuming any prior knowledge of condition (\bzdyo).
We start with the autonomous case, for constant $a_n$, for which there is a confined singularity with a (co-dimension 1) singular locus which corresponds to initial conditions $z_0, z_1, z_2$ and $z_3=a z_2^2/z_1$. The resulting singularity pattern is
$$\{{az_2^2\over z_1}, 0,0,*,*,*\},\eqdef\spthree$$
where $*$ again indicates a finite value where some (initial) degrees of freedom are missing. Only at the next step, i.e. at $z_9$, is the full complement of degrees of freedom recovered. The nature of this pattern is therefore quite similar to that of the confined pattern (\sptwo) for the third order map (\bztes).

When $a_n$ is not a constant, the singular locus for (\bzhex) is situated at $z_3=a_1z_2^2/z_1$ and calculating subsequent iterates we obtain essentially the same singularity pattern (\spthree) up to $z_8$. However, for general $a_n$ the next iterate, $z_9$, now diverges. The condition for the value of this iterate to be finite, condition (\bzdyo), in fact also allows for the recovery of all degrees of freedom. In this sense, this deautonomisation is actually more straightforward than that for the third order equation above. It is in fact a textbook example of a deautonomisation based on a confined pattern obtained from an autonomous map: without imposing any constraints on the coefficient functions in the nonautonomous map, undesired values appear in the singularity pattern and the constraints that prevent those values from appearing yield the correct deautonomisation of the map. In this sense the deautonomisation of (\bztes), while still being pretty standard, was actually slightly more subtle.

As was the case for the third order mapping (\bztes), an anticonfined singularity pattern does also exist here:
$$\{\cdots,\infty^9,\infty^7,\infty^6,\infty^5,\infty^3,\infty^2,\infty^2,\infty,\bol,\bol,\bol,0,0,\bol,\bol,\bol,\infty,\infty^2,\infty^2,\infty^3,\infty^5,\infty^6,\infty^7,\infty^9,\cdots\}.\eqdef\acptwo$$
Note that, due to the presence of three consecutive finite values in two different places, there are again two distinct ways to enter this anticonfined pattern: either through the initial condition $z_0,z_1,z_2,0$, or through $z_0,z_1,z_2,\infty$. 

The integrability of (\bzhex), subject to condition (\bzdyo), is confirmed by the study of the degree growth of the iterates. Starting from generic $z_0$, $z_1$ and $z_2$ we compute the degree of the iterates in terms of $z_3$ and find: 
0, 0, 0, 1, 2, 2, 3, 5, 7, 9, 11, 14, 17, 20, 24, 28, 32, 36, 41, 46, 51, 57, 63, 69, $\cdots$, again a quadratic degree growth.

\medskip
2.2 {\sl Calculation of anticonfined patterns and degree growth via max-plus algebra}
\smallskip
Let us pause the discussion here for a moment and remark that these anticonfined patterns can, in fact, be obtained in a much simpler way than by a mere brute force calculation, by using a max-plus analogue of the map. For example, the max-plus version of equation (\bztes),
$$Y_{n+1}=Y_{n-2}+\max(Y_n,Y_{n-1})-Y_{n-1},\eqdef\mpone$$
 is obtained by setting $y_n=e^{\delta Y_n}$ in  (\bztes) and then applying the operation $\lim_{\delta\to+\infty} {1\over\delta}\log (\,\cdot\,)$ while regarding all $a_n$ as negative and constant w.r.t. to this operation.  (People familiar with ultradiscrete limits will recognise here that what we are talking about is simply the {\sl ultradiscretisation}  [\refdef\toki] of the mapping). 

Taking initial conditions $Y_0=0, Y_1=0,Y_2=-1$, the max-plus equation (\mpone) yields  the sequence 0, 0, $-1$, 0, 1, 0, 0, 1, 1, 0, 1, 2, 1, 1, 2, 2, 1, 2, 3, 2, 2, 3, 3, 2, 3, 4, 3, 3 $\cdots$, which is exactly the sequence of multiplicities in the anticonfined pattern (\acpone) obtained from initial values $(y_0,y_1,0)$, if one identifies positive entries in the sequence with multiplicities of $\infty$ and negative ones with those for $0$. Iterating backwards from this initial condition, which can be done simply by considering the mapping $Y_{n-1}=Y_{n+2}-\max(Y_n,Y_{n+1})+Y_{n}$, we find the sequence $~\cdots$,  $-3$, $-2$, $-2$, $-3$, $-2$, $-1$, $-2$, $-2$, $-1$, $-1$, $-2$, $-1$, 0, $-1$, $-1$, 0, 0, $-1$,  in perfect agreement with the exponents of the zeros obtained by direct calculation of the anticonfined pattern. 

Note that this correspondence is rooted in the fact that the anticonfined pattern (\acpone) for mapping (\bztes) remains the same for any (non-zero) values for the parameters $a_n$, in which case one is allowed to assume $-a_n>0$ which is the condition needed to derive the max-plus version (\mpone). This fact also implies that the anticonfined pattern cannot contribute any information that could be useful in the deautonomisation of  the original mapping.

The same observation holds for the mapping (\bzhex). Its anticonfined pattern (\acptwo) is the same for any non-zero choice of the $a_n$ and, hence, one may assume suitable values for them such that one obtains the following max-plus version for equation (\bzhex):
$$Z_{n+1}=Z_{n-2}+Z_n+\max(Z_n+Z_{n-2}, 2Z_{n-1})-2Z_{n-1}-Z_{n-3}.\eqdef\mptwo$$
Starting from initial conditions $Z_0=0, Z_1=0, Z_2=0, Z_3=-1$, corresponding to initial conditions $z_0,z_1,z_2,0$ for (\bzhex), its max-plus version yields the sequence 0, 0, 0, $-1$, $-1$, 0, 0, 0, 1, 2, 2, 3, 5, 6, 7, 9, 11, 12, 14, 17, 19, 21, 24, 27, $\cdots$, which is exactly the sequence of multiplicities propagating towards the right in the anticonfined pattern (\acptwo). (The multiplicities emanating towards the left are the same as those on the right, due to the forward/backward symmetry of equation (\bzhex), which is of course inherited by its max-plus version).

Moreover, following the singularity balancing procedure we introduced in [\higher] based on Halburd's method [\hardrod],  it is clear that the degree growth of the mapping (\bzhex)  is given directly by the multiplicities in the infinities of the anticonfined pattern, since the confined singularity (\spthree) does not  contribute any infinities at all. (Note that this is not the case for mapping (\bztes)). More precisely, as there are two separate instances of the
 anticonfined pattern (\acptwo) to take into account, we have in fact that the degree (calculated in $z_3$ for generic values of $z_0, z_1$ and $z_2$) is given by $d_n=Z_{n+5}+\max(Z_{n},0)$, where $Z_n$ are the multiplicities obtained from the max-plus equation (\mptwo) for the initial condition (0, 0, 0, $-1$). We find $7d_n=n^2-2n-1+\phi_7(n)$, (where $\phi_7(n)$ is a period-7 function obtained the repetition of the string $[1,2,1,-2,0,0,-2]$), which reproduces the degrees obtained by direct calculation for $n\geq5$. 

\medskip
2.3 {\sl Scenario II -- Deautonomisation based on an unconfined pattern: appearance of undesired values}
\smallskip
As explained above, up to now we only encountered classical cases of deautonomisations where a confined singularity pattern is preserved. In this section we shall give two examples of higher order mappings that do not have confined singularities in co-dimension 1, but which can still be deautonomised based on information gained from their non-confined patterns.

A first example is obtained by coupling equation (\orgeq) to a linearisable one. For this we choose a simple form for a projective mapping and introduce the coupling
$$u_{n+1}={x_n\over u_n u_{n-1}},\eqdef\bdtri$$
resulting into the fourth order mapping
$$u_{n+1}={u_nu_{n-1}u_{n-2}-a_{n-1}\over u_n^2u_{n-1}^3u_{n-2}^2u_{n-3}}.\eqdef\bdtes$$
This mapping does not have any confined or anticonfined singularities, only unconfined ones. Starting from generic $u_0$, $u_1$, $u_2$ and $u_3=a/(u_1u_2)$ we obtain (for constant $a\neq0$) the singularity pattern 
$$\{{a\over u_1u_2}, 0,\infty^2,\bol, 0^3,\infty^3,\bol, 0^3,\infty^3,\bol, 0^3,\infty^3,\cdots\},\eqdef\IIprojucpone$$
where the string $[\bol, 0^3,\infty^3]$ repeats indefinitely, leading to an {\sl unconfined} singularity.  

When we consider the non-autonomous case, entering the singularity through $u_3=a_2/(u_1u_2)$, we obtain a singularity pattern that is different from that for the autonomous case: 
$$\{{a_2\over u_1u_2}, 0,\infty^2,\bol, 0^3,\infty^3,\infty^1,0^4,\infty^2,\infty^3,0^4,\bol,\cdots\},\eqdef\IIprojucptwo$$
in which undesired 
values that do not occur in the pattern (\IIprojucpone), arise as of $u_9$. Requiring that the two patterns be identical leads to a single constraint, obtained at the level of $u_9$, which is precisely (\bzdyo).

Imposing the constraint (\bzdyo) in fact makes the mapping (\bdtes) integrable. Computing its degree growth in terms of $u_3$ (for generic values of $u_0, u_1$ and $u_2$) gives the sequence 0, 0, 0, 1, 2, 2, 5, 7, 10, 15, 18, 27, 34, 41, 56, 66, 81, 100, 116,$\cdots$, which appears to be cubic. To generate  a longer sequence of  degrees we can use the method we introduced in [\fast] but which requires knowledge of all the singularity patterns in co-dimension 1 for this mapping. As we mentioned earlier, in co-dimension 1, the mapping (\bdtes) does not have any confined or anticonfined singularities but, besides the unconfined pattern discussed above, it possesses two singularities that are part of cyclic singularity patterns. The first pattern arises from a value 0,
$$\{\bol,\bol,\bol,0^1,\infty^2\bol,0^3,\infty^3,\infty^1,0^4,\infty^2,\infty^3, 0^4,\bol,\infty^4,0^3,0^1\infty^3,0^1,0^1\infty^1\},\eqdef\IIcyclpatone$$
and the second one from an $\infty$: 
$$\{ \bol,\bol,\bol,\infty^1,0^1,0^1,\infty^3,0^1,0^3,\infty^4,\bol, 0^4,\infty^3,\infty^2,0^4,\infty^1,\infty^3,0^3\bol,\infty^2,0^1\}.\eqdef\IIcyclpattwo$$

Based on the unconfined pattern (\IIprojucpone) and the two cyclic patterns, (\IIcyclpatone) and (\IIcyclpattwo), we can apply the algorithm we proposed in [\fast] and find for the first 100 degrees:

0, 0, 0, 1, 2, 2, 5, 7, 10, 15, 18, 27, 34, 41, 56, 66, 81, 100, 116, 139, 163, 189, 219, 252, 289, 329, 371, 419, 469, 523, 582, 642, 711, 781, 854, 938, 1020, 1110, 1207, 1304, 1411, 1522, 1638, 1761, 1890, 2026, 2168, 2315, 2471, 2632, 2800, 2976, 3156, 3348, 3544, 3746, 3962, 4179, 4407, 4645, 4886, 5140, 5401, 5670, 5949, 6237, 6535, 6842, 7157, 7484, 7819, 8164, 8520, 8883, 9261, 9646, 10040, 10451, 10866, 11295, 11737, 12185, 12649, 13123, 13608, 14106, 14616, 15139, 15674, 16220, 16781, 17353, 17938, 18537, 19146, 19773, 20410, 21059, 21728, 22404, 23097. 

Once a sufficiently long degree sequence is available, it is straightforward to obtain an explicit expression that  represents it faithfully. We find thus 
$$d_n={1\over42}n^3-{1\over14}n^2+{38\over21}+ \phi_{21}(n),\eqdef\bdtesmiso$$
 where $\phi_{21}(n)$ is the periodic function obtained by repetition of the string $[-37, -36, -17,- 4, -21, 13, 11, $ $12, 34, -10, 45, 28, -22, 60, -2, -1, 39, -32, -7,$ $ -15, -38]$, where each entry is to be divided by 21.

The above deautonomisation, although performed on an unconfined singularity pattern, does not differ all that much from standard deautonomisation (based on confined patterns): parameter conditions are obtained at iterates where 
undesired values arise in the pattern for the nonautonomous mapping and once those values have been removed the patterns for the autonomous and nonautonomous cases coincide. 

However, more subtle scenarios are possible. Consider for example the fifth order mapping obtained from coupling equation (\bzhex) to one more linear equation: 
$$w_{n+1}=z_nw_n.\eqdef\bzhep$$
We find for $w$ the equation
$$w_{n+1}={w_n^2w_{n-2}w_{n-4}\over w_{n-1}^4w_{n-3}^3}\left(w_nw_{n-2}^3-a_{n-3}w_{n-1}^3w_{n-3}\right).\eqdef\bzoct$$
In the autonomous case, i.e. for $a_n=a$ constant, one can enter a singularity (in co-dimension 1) through $w_4=a w_1w_3^3/w_2^3$. Starting from generic $w_0$, $w_1$, $w_2$ and $w_3$ we find that this singularity leads to an  unconfined singularity pattern:
$$\{{aw_1w_3^3\over w_2^3}, 0^1,0^2,0^2,0^2,0^2,0^2,0^2,\cdots\}.\eqdef\ucpatone$$

In the non-autonomous case, starting from $w_4=a_1w_1w_3^3/w_2^3$, we obtain a different unconfined singularity pattern
$$\{{a_1w_1w_3^3\over w_2^3}, 0^1,0^2,0^2,0^2,0^2,0^1, \infty^1, \infty^3, \infty^6, \cdots\}\eqdef\ucpattwo$$
in which, in particular, $w_{10}$
is a simple zero rather than a double one as it would have been if the $a_n$ were constant. This change in the multiplicity of $0$ gives rise to a sequence of undesired 
entries (in comparison with the pattern that one would like to preserve): infinities that appear with increasing multiplicties.  Requiring that $w_{10}$ instead has a double zero, leads to the condition (\bzdyo) and implementing this condition reverts the entire singularity pattern to that of the autonomous case. Thus, just as in the previous case, the deautonomisation condition for the (integrable) mapping (\bzoct)  is that its unconfined singularity pattern be preserved, and here this is actually achieved most easily by comparing multiplicities in the patterns rather than by looking at the 
undesired values that appear further along in the pattern. 

The presence of this unconfined pattern is reflected in the degree growth for the mapping (\bzoct).  Computing the  degree of its iterates in terms of $w_4$, with $w_0, w_1,w_2$ and $w_3$ generic,  leads to the sequence: 0, 0, 0, 1, 3, 5, 8, 13, 19, 27, 38, 52, 69, 89, 113, 141, 173, 209, 250, 296, 347,$\cdots$. This growth is cubic and the sequence of degrees can be, for $n>6$, represented by the expression 
$$d_n={1\over24}n^3+{3\over16}n^2-{47\over12}n+{539\over32}+\phi_4(n),\eqdef\bzenn$$
where $\phi_4(n)$ is
obtained by the repetition of the string $[{-5\over32},{-3\over32},{3\over32},{5\over32}]$.

Moreover, two anticonfined singularities exist as well for this mapping:
$$\scalebox{0.9}{ $\{\cdots,0^{33},0^{24},0^{17},0^{11},0^6,0^3,0,\infty^1,\infty^2,\infty^2,\infty^2,\infty^2,\infty,w_0,w_1,w_2,w_3,\infty,\infty^3,\infty^5,\infty^8,\infty^{13},\infty^{19},\infty^{26},\infty^{35},\cdots\}$}$$
and 
$$\scalebox{0.9}{$\{\cdots,0^{129},0^{108},0^{89},0^{72},0^{58},0^{46},0^{35},0^{26},0^{19},0^{13},0^{8},0^5,0^3,0,w_0,w_1,w_2,w_3,0,0^2,0^2,0^2,0^2,0,\infty,\infty^3,\infty^6,\infty^{11},\cdots\}$},$$
which are related to each other by exchanging zeros and infinities, as well as the evolution directions. Note that, unlike the unconfined pattern (\ucpattwo), these patterns do not depend on the detailed values $a_n$ takes (as long as these are non-zero) and that the multiplicities of the infinities in the two patterns grow cubically with $n$. Given that the unconfined  singularity pattern (\ucpatone) of the mapping (\bzoct) does not contain any infinities, the degree $d_n$ of its iterates (in terms of $w_4$, with $w_0, w_1,w_2$ and $w_3$ generic)  is in fact given by the sum of the multiplicities of the infinities in the two anticonfined patterns,
$$d_n=W_n+\max(\widetilde W_n,0),\eqdef\dnforw$$
where $W_n$ and $\widetilde W_n$ are obtained from the max-plus version of (\bzoct), with initial conditions $W_0=0, W_1=0, W_2=0,W_3=0$ and with  $W_4$ ($\widetilde W_4$) equal to 1 (and $-1$) respectively. This confirms that the nonautonomous map (\bzoct), subject to condition (\bzdyo), has the same (cubic) degree growth as the autonomous one and that this deautonomisation therefore not only preserves the unconfined singularity pattern (\ucpatone) of the mapping, but also its integrable character.

Although the multiplicities of the singular entries in the singularity patterns (\ucpatone) and (\ucpattwo) obviously yield crucial information on the dynamics of this mapping, its deautonomisation could have been carried out without taking into account the change of multiplicity in the $w_{10}$ iterate: the appearance of undesired values as of $w_{11}$ does actually, in itself, give enough information to successfully deautonomise the mapping.
The deautonomisation procedure becomes more subtle, however, when such undesired  values do not appear.

\medskip
2.4 {\sl Scenario III -- Deautonomisation based on an unconfined pattern: changes in the multiplicities}
\smallskip
In order to obtain an example of a more challenging deautonomisation procedure, let us couple the integrable mapping (\bzena) to an affine one,
$$y_{n+1}=x_ny_n+b_{n+1},\eqdef\bzdek$$
where we suppose that the $b_n$ are non-zero. The resulting third order equation is now
$$y_{n+1}=y_{n-2}y_n{y_n-b_n-a_{n-1}y_{n-1}\over(y_n-b_n)(y_{n-1}-b_{n-1})}+b_{n+1}.\eqdef\bdena$$
Several co-dimension 1 singularities exist. Starting from generic $y_0, y_1$ and $y_2=0$ we obtain a confined singularity pattern
$$\{0,b_3,*\},\eqdef\IIIcpat$$ 
where at the (finite) value for $y_4$ one has not yet regained the full freedom of the initial conditions $y_0, y_1$ and, hence, it is to be considered part of the singularity pattern. However, both initial conditions $y_0$ and $y_1$ are automatically recovered at the level of $y_5$, resulting in the confined pattern (\IIIcpat) of length three for any value of $b_n$. Hence, this confined pattern does not yield any useful information vis-a-vis the deautonomisation of the mapping.

 Next we consider the singularity that is induced by taking $y_2= b_2+a_1y_1$ and we start by examining the autonomous case $b_n=b\neq0, a_n=a\neq0$, for all $n$, for which we obtain the following unconfined pattern: 
 $$\{b+ay_1,b,\infty^1,\infty^2,\infty^1,\infty^1,\infty^1,\infty^1,\infty^1,\cdots\}.\eqdef\IIIucpatone$$
If, on the other hand, we consider the non-autonomous case we obtain the singularity pattern
 $$\{b_2+a_1y_1,b_3,\infty^1,\infty^2,\infty^1,\infty^1,\infty^2,\infty^2,\infty^1,\infty^3,\infty^3\cdots\},\eqdef\IIIucpattwo$$
 where again the singularity is not confined. We remark that $y_8$ diverges as $\infty^2$, instead of $\infty^1$ in the autonomous case, and that subsequent multiplicities differ considerably, but that the pattern (\IIIucpattwo) does not contain any  undesired  values compared to the autonomous one. However, requiring that we obtain the same multiplicity at $y_8$ for the non-autonomous and autonomous cases leads to a constraint on the $a_n$ which is precisely ($\bzdyo$). In fact, if this condition is imposed the singularity pattern in the non-autonomous case becomes identical to (\IIIucpatone), with a $\infty^1$ divergence for all $y_n$ with $n>5$. Which begs the question: what about $b_n$? It turns out that $b_n$ is, essentially, a completely free function, a situation which is not uncommon for linearisable (or, as in the present case, partially linearisable) systems [\refdef\limproc]. 

Anticonfined patterns also exist for (\bdena). Entering the singularity with $y_2=b_2$ we find the pattern 
$$\{\cdots,\infty^1,\infty^1,0,\bol,\infty^1,\bol,0,\bol,\bol,b_2,\infty^1,\infty^2,\infty^1,\infty^1,\infty^2,\infty^2,\infty^1, \infty^2 \cdots\},\eqdef\IIIacpone$$
while entering through an infinity we obtain 
$$\{\cdots,\bol,\infty^1,\infty^1,0,\bol,\infty^1,\bol,\bol,\infty^1,\infty^1, \bol,\infty^1,\infty^2,\infty^1,\infty^1,\infty^2,\infty^2,\infty^3 \cdots\},\eqdef\acptwo$$
for any functions $a_n$ or $b_n$. 

Under condition ($\bzdyo$) and for free functions $b_n$, the degree growth of the mapping (\bdena) computed in terms of $y_2$, with generic $y_0$ and $y_1$, is 0, 0, 1, 2, 3, 5, 8, 12, 17, 23, 31, 40, 51, 64, 79, 96, 115, 137, 161, $\cdots$, which is once more a cubic growth. It can be represented, for $n>1$, by the expression
$$d_n={1\over42}n^3+{1\over14}n^2-{2\over21}n+{6\over7}+\phi_7(n),\eqdef\bdyo$$
where $\phi_7(n)$ is the period-7 function obtained by repetition of the string $[{1\over7},{-1\over7},{1\over7},-{1\over7},-{1\over7},0,{1\over7}]$.

Traditionally, when testing a mapping for the singularity confinement property, one might not feel the need to take into account the multiplicities with which singular values appear in the singularity patterns. As we explained in section 2, knowledge of these multiplicities is crucial if one wants to calculate the degree growth of a given mapping based on its singularity patterns [\hardrod, \express, \fast]. The above scenario however shows that when it comes to deautonomising higher order mappings, it is imperative that one take into account the multiplicities of the singular values because without that information, in some cases, the singularity patterns do not yield enough information to deautonomise the mapping.
\bigskip
3. {\scap Higher order non-integrable maps}
\medskip
Up to now we have studied different scenarios that arise when deautonomising higher order integrable maps. The main difference with the second order case being that such mappings very often exhibit non-confined singularities. Luckily enough, as we have shown, such singularities can be used to obtain integrable deautonomisations if one extends the singularity confinement analysis to include information on the multiplicties of the singular values.

We shall now look at the deautonomisation of some nonintegrable higher order mappings, but first we should explain what we mean by `deautonomisation' for such mappings. As is clear from the classification of birational second order mappings by Diller and Favre [\dilfav], although quite rare, it is possible for a non-integrable mapping to possess a space of initial conditions. In that case a deautonomisation procedure can be defined rigorously, in algebro-geometric terms (see e.g. [\refdef\takenawahiv, \alex]), which is such that the deautonomised mapping is guaranteed to have the same dynamical degree as the original autonomous one. The first instance of such a deautonomisation was given by Takenawa [\takenawahiv] for the example of a confining non-integrable map found by Hietarinta and Viallet [\refdef\hiv].
In the following we shall look at two higher order non-integrable maps that are constructed by coupling the Hietarinta-Viallet map to different linear maps. Although both those maps have non-confined singularities (in co-dimension one) and can therefore not be `lifted' to a pseudo-isomorphism (contrary to the higher order mappings described in [\refdef\takcarst, \refdef\alextakcarst]), we will show that deautonomisations that preserve the singularity patterns of the original autonomous mapping---including multiplicities---result in nonautonomous mappings with the same dynamical degree as the original autonomous mapping. This result also suggests that the `full-deautonomisation' method we introduced to obtain the dynamical degree for autonomous second order maps, by finding judicious deautonomisations for them, might also work for higher order maps.

Finally, we end this section with a short discussion of nonautonomous mappings obtained from `late confinement' for a higher order integrable map, based on an unconfined singularity pattern for the latter, pushing in a sense our deautonomisation approach to its very  limits.

\medskip
3.1 {\sl Deautonomisation of higher order non-integrable maps}
\smallskip
We start with the standard example of the mapping introduced by Hietarinta and Viallet [\hiv], and consider its extension which, as we showed in [\redemption], corresponds to a full deautonomisation of that mapping with coefficients that reflect its dynamical degree:
$$x_{n+1}+x_{n-1}=x_n+{1\over x_n^2}+{f_n\over x_n}.\eqdef\bdpen$$
In the autonomous case where $f_n$ is constant (and equal to 0 in the `standard' version), the mapping has a confined singularity pattern: 
$$\{0,\infty^2,\infty^2,0\}.\eqdef\HiVpat$$ 
Requiring that this pattern remain the same in the non-autonomous case, we obtain for $f_n$ the constraint
$$f_{n+3}-2f_{n+2}-2f_{n+1}+f_n=0.\eqdef\bdhex$$
The largest root of the characteristic polynomial for this linear recurrence, $(\lambda^2-3\lambda+1)(\lambda+1)=0$, is $(3+\sqrt{5})/2\approx2.6180$, which is precisely the dynamical degree of the mapping [\redemption,\alex]. 

As explained in detail in  [\refdef\late], it is possible to postpone confinement for the mapping (\bdpen) to a later stage: a first possibility for such a late confinement arises at the end of the pattern
$$\{0, \infty^2, \infty^2,0,\infty,\infty,0, \infty^2, \infty^2,0\},\eqdef\HiVlongpat$$
and subsequent confined patterns are obtained by appending the string
$$[\infty,\infty,0, \infty^2, \infty^2,0],\eqdef\string$$
repeatedly to (\HiVlongpat). Moreover, confinement can be postponed indefinitely [\late], resulting in the unconfined singularity pattern that one would find for (\bdpen) for arbitrary functions $f_n$. The dynamical degree for this `generic' case of (\bdpen) is given by the largest root of the polynomial $\lambda^5-2\lambda^4-2\lambda^3+\lambda^2-\lambda-1$ associated to (\string): $\lambda_*\approx 2.67871$.

We now couple (\bdpen) to the linear mapping $y_{n+1}=x_ny_n$ and obtain the third-order mapping
$$y_{n+1}={y_n^3+y_{n-1}^3+f_ny_ny_{n-1}^2\over y_ny_{n-1}}-{y_ny_{n-1}\over y_{n-2}}.\eqdef\bdhep$$
In the autonomous case, (\bdhep) has the unconfined singularity pattern 
$$\{0,\infty,\infty^3,\infty^2,\infty^2,\infty^2,\infty^2,\cdots\}.\eqdef\HiVFCpat$$
However, when we consider a generic non-autonomous case, without imposing any conditions on the coefficients $f_n$,  we obtain the pattern 
$$\{0,\infty,\infty^3,\infty^2,\infty^3,\infty^4,\infty^3\cdots\}.\eqdef\HiVFCnonpat$$
For the two patterns to be  identical it suffices to ensure that the second $\infty^3$ in the latter pattern be  just a $\infty^2$ (cf. Scenario III in section 2.4). It turns out that the condition for this to happen is precisely (\bdhex), which also happens to give the dynamical degree for the nonautonomous mapping (which we verified using the Diophantine procedure). 

The coupled mapping (\bdhep)  has, as expected, also an anticonfined singularity pattern,
$$\{\cdots, 0^4,0^3,0^2,0^2,0^1,\bol,\bol,\infty^1,\infty^2,\infty^2,\infty^3,\infty^4,\infty^4,\infty^5,\cdots\},\eqdef\HiVFCac$$
which together with the unconfined pattern (\HiVFCpat) can be used to verify the dynamical degree (for example using the method introduced in [\fast]).

Next we consider the coupling of (\bdpen) to the affine mapping (\bzdek), $y_{n+1}=x_ny_n+b_{n+1}$, leading to
$$y_{n+1}={(y_n-b_n)^3+y_{n-1}^3-f_{n-1}y_{n-1}^2(y_n-b_n)\over y_{n-1}(y_n-b_n)^2}-{y_n(y_{n-1}-b_{n-1})-b_{n+1}y_{n-2}\over y_ny_{n-2}}.\eqdef\bdoct$$
We start by considering the case where $f_n$ is constant but we do not constrain $b_n$, for which we find there exists a confined singularity pattern. Starting from generic $y_0$, $y_1$ and $y_2=0$ we obtain 
$$\{0,b_3,*\}, \eqdef\HiVSCcp$$
where $y_4$ is finite but one has only recovered one degree of freedom. The remaining degree of freedom is recovered automatically at the next step, i.e. at $y_5$, and thus the singularity is indeed confined. 
However a non-confined pattern does also exist: if we start with $y_2=b_2$ we find 
$$\{b_2,\infty^2,\infty^4,\infty^3,\infty^3,\infty^3,\cdots\}.\eqdef\HiVSCnc$$

We turn now to the fully non-autonomous case where $f_n$ is also a function of $n$. While nothing changes for the confined pattern, the non-confined one becomes 
$$\{b_2,\infty^2,\infty^4,\infty^3,\infty^4,\infty^5,\cdots\}.\eqdef\HiVSCncnon$$
Requiring that the two patterns be identical yields the condition (\bdhex) for the function $f_n$, which as we have seen grows asymptotically as powers of $(3+\sqrt{5})/2$, which was checked to correspond to the value of the dynamical degree using the Diophantine approach. 

Again, the precise form of the functions $b_n$ does not influence the singularity patterns, nor does it have any influence on the value of the dynamical degree.

Note that the value we found for the dynamical degrees of (\bdpen), (\bdhep) or (\bdoct), which are all the same, is somewhat smaller than the dynamical degree ($2.67871...$) we would find for these mappings for arbitrary functions $f_n$ (which is again the same  for all three mappings since the linear couplings we considered cannot change the dynamical degree in these non-integrable cases).

\medskip
3.2 {\sl A case of `late confinement' based on an unconfined pattern}
\smallskip
As explained in the introduction, delaying the application of the confinement constraints leads to non-integrable mappings. Consider for instance the mapping 
$$x_{n+1}+x_{n-1}={a_n\over x_n}+{1\over x_n^2}.\eqdef\bdenn$$
In the autonomous case it has a confined singularity pattern $\{0,\infty^2,0\}$ which means that in the non-autonomous case, starting from a generic $x_0$ and $x_1=0$, requiring that $x_4$ depend on $x_0$, leads to a confinement constraint on $a_n$.  (The latter is $a_{n+1}-2a_n+a_{n-1}=0$, i.e. $a_n$ is linear in $n$, and the mapping is a well-known discrete Painlev\'e equation). However, it is possible to ignore this possibility to confine the singularity. It turns out that another opportunity appears after four more iterations (and, in fact, infinitely many such opportunities appear periodically, every four iteration steps). Focusing on the first `late' confinement opportunity, we obtain the (confined) singularity pattern 
$$\{0,\infty^2,0,\infty,0,\infty^2,0\},\eqdef\latconfpat$$
and the associated constraint
$$a_{n+5}-2a_{n+4}+a_{n+3}-a_{n+2}+a_{n+1}-2a_n+a_{n-1}=0.\eqdef\bddek$$
The largest root of the characteristic polynomial of (\bddek) is $\lambda={1+\sqrt{13}\over 4}+\sqrt{{\sqrt{13}-1\over 8}}$, with numerical value 1.7220838$\dots$, which is precisely the dynamical degree of the mapping when $a_n$ obeys (\bddek).

Let us now consider the triple coupling of (\bdenn) to  the linear mapping (\bztri). In order to avoid lengthy expressions we write is as a system composed from (\bdenn) and 
$$w_{n+1}=x_{n-2}{w_{n-2}w_n^3\over w_{n-1}^3}.\eqdef\bvena$$
We start, as for (\bdenn), with $x_0, x_1=0$ and with finite values for $w_0,w_1$ and $w_2$ and obtain $w_3=x_0 w_0 w_1^{-3} w_2^3$ and $w_4=w_5=w_6=0^1$. In the autonomous case, when $a_n$ is constant, we find that for all $n\geq7$, $w_n=0^1$ as well and, thus, that the singularity is unconfined.  However, in the nonautonomous case we find that $w_7$ takes a finite non-zero value and that as of $n\geq8$ the values in the singularity pattern change dramatically:  $\infty^1,\infty^4,\infty^8,\infty^{14},\infty^{21}, ...$. Given the increasingly fast growth in the multiplicities this would imply that the coupled mapping is non-integrable.

Requiring that $w_7$ be $0^1$ leads to the constraint $a_{3}-2a_2+a_{1}=0$ and the singularity pattern (in the $w$ variable) reverts to that for the autonomous one:
$$\{\bol,\bol,0^1,0^1,0^1,0^1, \cdots\}.\eqdef\latucpatone$$
However, we may choose to ignore this possibility and try to curb the growth in the multiplicities of the singular values at a subsequent iterate. It turns out that such a possibility appears at the level of $w_{11}$ and it corresponds precisely the late confinement condition (\bddek). For example, when this condition is implemented, the multiplicities of $\infty$ in $w_{11}$ and $w_{12}$ are 13 and 19 instead of 14 and 21 and similarly for the subsequent iterations.  Moreover, the dynamical degree of the coupled mapping, computed through the Diophantine approach, is indeed approximately 1.722, as predicted by the late confinement condition. 

Delaying further the application of a condition that would (somewhat) curb the growth is always possible. As explained in [\late], it is possible to keep postponing such a condition indefinitely. We shall not go into these details here but just give the characteristic polynomial  for an infinitely-late application of such a condition: 
$$k^6-2k^5+k^4-k^3+k-2 = 0.\eqdef\lastpoly$$
The largest root of the polynomial (\lastpoly) is (approximately)
1.767, which is the dynamical degree of a generic nonautonomous version of the coupled mapping (\bdenn)$\cdot$(\bvena), a value again confirmed by a Diophantine-based calculation.
\bigskip
4. {\scap Conclusions}
\medskip
In this paper we have explored possible deautonomisation scenarios for higher order mappings, integrable as well as non-integrable, based on their singularity structure. As nonconfined singularities are prevalent at higher order---even in the case of integrable mappings---we have
 focused primarily on the deautonomisation of mappings that were constructed through coupling with linear (or linearisable) maps. This was done in order to have sufficiently many interesting examples to illustrate every deautonomisation scenario we could imagine at present. We do not wish to claim that our list of scenarios is exhaustive---we will come back to this point at the very end of these conclusions.

The main lesson to be drawn from the examples we presented here is that singularity analysis can indeed be used to obtain nonautonomous versions of a given higher order map, such that the nonautonomous map has the same dynamical degree as the original autonomous one. For integrable maps this means that the deautonomisation preserves integrability, which is a very strong property. For non-integrable maps the statement that our deautonomisations preserve the dynamical degree might appear much less interesting but we would beg to differ. In the cases with exponential degree growth that we presented, the dynamical degree was sometimes
considerably smaller than what one would expect for a generic map of the same form. The exact cancellations,
 that are necessary in the successive iterates of the map to have such slower growth, are highly intricate and cannot be obtained just haphazardly. Moreover, at least in the cases at hand, our method yields non-autonomous mappings the coefficients of which still reflect the value of the dynamical degree of the map, just as they do in the full-deautonomisation procedure for second order mappings. It would be extremely interesting to understand the geometric reasons why our deautonomisation procedure has this particular property, especially in the case where the deautonomisation was performed on unconfined patterns. Unfortunately, at present, it seems that the tools required for a thorough algebro-geometric understanding of such deautonomisations are still lacking, even in the integrable case.

Another important conclusion we can draw from our analysis is that, in order to deautonomise a given map successfully, it is imperative that one include the information on the 
multiplicities 
with which singular values appear in the singularity analysis. We have demonstrated that, without this information, there are cases where the pattern itself does not yield enough information to deautonomize. Moreover, in some cases, as shown at the end of section 3.2, the information on the multiplicities of the singular values might even be used to construct the equivalent of ``late confinements for non-confining maps". As the previous sentence amply demonstrates, with that statement we are most likely both linguistically as well as mathematically at the very edge up to which the method can be pushed.

We would like to end this paper however with another, different, example of a possible limitation to the method we presented. As explained at the very beginning, we limited our singularity analysis to singularities that arise in co-dimension 1. The main reason being that it seems that for higher order  mappings, the best analogy one has to the notion of a space of initial conditions, is to try to regularize a mapping, through successive blow-ups, to a pseudo-isomorphism  [\bedkim, \takcarst], i.e. an isomorphism in co-dimension 1. However, in order to deautonomize higher order mappings this notion might not be sufficient, not even in the case of integrable mappings. Consider for example the third order mapping given by the equation
$$y_{n+1} = 1 + {a\over y_n + y_{n-1} } - 2 y_n -2 y_{n-1} - y_{n-2},\eqdef\badNV$$
where $a$ is a non-zero constant. As shown in [\refdef\joshiviallet], this mapping has cubic degree growth and is actually obtained by coupling the simple QRT map
$$x_{n+1}+x_n+x_{n-1}=1+{a\over x_n},\eqdef\standPIaut$$
to the linear equation $x_n=y_n+y_{n+1}$. In co-dimension 1, the map (\badNV) has a particularly simple singularity structure. Starting from $y_0, y_1$ generic and $y_2=\infty$ or $y_0, y_2$ generic and $y_1=\infty$ simply yields cyclic singularity patterns (with period 6) and the case $y_2+y_1=0$ (for generic $y_0$) yields the unconfined singularity pattern:
$$\{-y_1, \infty^1, \infty^1, \infty^1, \infty^1, \infty^1, \cdots\}.\eqdef\badpat$$
Since cyclic patterns yield no information at all vis-a-vis the deautonomisation procedure, one has only the unconfined pattern (\badpat) to work with if one would like to deautonomise (\badNV) by changing the constant $a$ to some well-chosen function $a_n$, such that the mapping stays integrable. Interestingly, it turns out that the pattern (\badpat), including the multiplicities in it, remains the same for any choice of $a_n$. This despite nonautonomous versions of (\badNV) exhibiting exponential growth when $a_n$ is chosen arbitrarily. (Note that this is not a contradiction: as the singularity patterns for the mapping only contain one single value, $\infty$, Halburd's method [\hardrod, \express] for obtaining degree growths from the singularity patterns does not apply and there is no direct relation between the multiplicities in the patterns and the growth of the degrees).

The singularity analysis of mapping (\badNV) does give an important clue however, as to what exactly is going on in this case. For generic $y_0$ and $y_1$, calculating successive iterates of $y_2=-y_1+\epsilon$ we obtain $y_3= a\epsilon^{-1}$ and then $y_n= (-1)^{n+1} 2 a\epsilon^{-1}$ for all $n\geq4$, which is what is represented in the unconfined pattern (\badpat). For non-constant $a_n$ however, we find that $y_3=a_1 \epsilon^{-1}$ and $y_4=-2 a_1\epsilon^{-1}$ and $y_5=2 a_1\epsilon^{-1}$, but that the subsequent pattern of alternating signs of the residues in $\epsilon$ is broken. In particular we find $y_6= a_1(a_1-a_2-a_3-a_4)/(a_2-a_1+a_3) \epsilon^{-1}, y_7= 2a_1 a_4/(a_2-a_1+a_3)\epsilon^{-1}$ and $y_8=-y_7$ for the next few iterates. Requiring that the function $a_n$ is such that $y_6=-y_5$, for general $n$, amounts to the condition
$$a_{n+3} - a_{n+2} - a_{n+1} + a_n=0,\eqdef\dPIcond$$
which also implies that $y_8=-y_7=y_6=-y_5=y_4$, just as in the autonomous case. This condition is of course nothing but the constraint that appears in the deautonomisation of the QRT map (\standPIaut) to the standard discrete P$_{\rm I}$ equation [\singconf]. Hence, under this constraint the nonautonomous version of the mapping (\badNV) becomes an integrable one (since it is equivalent to the coupling of a discrete Painlev\'e to a linear equation). In particular, we find that the alternating sign in the residues in $\epsilon$ that we observe in the singularity analysis of the singularity induced by $y_2=-y_1+\epsilon$, is by itself sufficient to reduce the degree growth from an exponential one (for arbitrary $a_n\neq 0$) to a cubic one. This suggests that the indefiniteness in the mapping (\badNV) that gives rise to the extraordinary cancellations that are needed to reduce exponential to polynomial degree growth, are not only due to all  successive iterates being infinite, but also to the fact that two such successive infinities cancel out when summing $y_n$ and $y_{n-1}$ (as of $n\geq5$). Thus it seems that these cancellations are arising at a deeper level that regular singularity analysis does not capture. 

Actually, from the shape of the singularity pattern (\badpat), it is clear that the residues in $\epsilon$ that appear in the `raw' version that comes out of the singularity analysis, can be regarded as homogeneous coordinates for points on the plane at infinity in $\Bbb P^3$. Introducing homogeneous coordinates, for example, as $y_{n-2}=Y_1/Y_0, y_{n-1}=Y_2/Y_0$ and $y_n=Y_3/Y_0$, we obtain the following mapping on $\Bbb P^3$,
$$[Y_0:Y_1:Y_2:Y_3] \mapsto [Y_0 (Y_2+Y_3):Y_2 (Y_2+Y_3):Y_3 (Y_2+Y_3):a Y_0^2 + (Y_2+Y_3)(Y_0-Y_1)-2(Y_2+Y_3)^2],\eqdef\Pthreemap$$
which is singular at $[0:Y_1:Y_2:-Y_2]$ (and only at this locus). We shall not go into any further details here, but this singularity leads to an unconfined singularity pattern in the autonomous case, and preserving this pattern upon deautonomisation leads to the condition (\dPIcond). The problem that arises here, however, is that this singular locus for the mapping (\Pthreemap) has co-dimension 2, which contradicts our basic ansatz to restrict our singularity analysis to co-dimension 1 singular loci. Hence, it would seem that when deautonomising some mappings, for example those with unconfined singularity patterns with homogeneous multiplicities for a single singular value, as for mapping (\badNV), a more refined analysis that takes into account singular loci in higher co-dimensions than 1, is required.
\bigskip
{\scap Acknowledgements}
\medskip
RW would like to thank the Japan Society for the Promotion of Science (JSPS)
for financial support through the KAKENHI grant 23K22401.

\bigskip
{\scap References}
\medskip
\item{[\singconf]} B. Grammaticos, A. Ramani and V. Papageorgiou, {\sl Do integrable mappings have the Painlev\'e property?}, Phys. Rev. Lett.{ \bf67} (1991) 1825-1828.
\item{[\qrt]} G.R.W. Quispel, J.A.G. Roberts and C.J. Thompson, {\sl Integrable mappings and soliton equations II}, Physica D {\bf 34} (1989) 183-192.
 \item{[\redemption]} A. Ramani, B. Grammaticos, R. Willox, T. Mase and M. Kanki, {\sl The redemption of singularity confinement}, J. Phys. A {\bf 48} (2015) 11FT02.
 \item{[\alex]} A. Stokes, T. Mase, R. Willox and B. Grammaticos, {\sl Deautonomisation by Singularity Confinement and Degree Growth}, J Geom Anal {\bf 35} (2025) 65.
 \item{[\dilfav]} J. Diller and C. Favre, {\sl Dynamics of bimeromorphic maps of surfaces}, Amer. J. Math. {\bf123} (2001) 1135-1169.
 \item{[\maseth]} T. Mase, {\sl Studies on spaces of initial conditions for non-autonomous mappings of the plane}, Journal of Integrable Systems {\bf 3} (2018)  xyy010.
\item{[\royal]} T. Mase, R. Willox, B. Grammaticos and A. Ramani, {\sl Deautonomisation by singularity confinement: an algebro-geometric justification}, Proc. Roy. Soc. A {\bf471} (2015) 20140956.
\item{[\hivtwo]} J. Hietarinta and C-M. Viallet, {\sl Discrete Painlev\'e I and singularity confinement in projective space}, Chaos, Solitons and Fractals {\bf11}  (2000) 29-32.
\item{[\blancdeserti]} J. Blanc and J. D\'eserti, {\sl Degree growth of birational maps of the plane}, Ann. Sc. Norm. Super. Pisa Cl. Sci. {\bf XIV} (2015) 507-533.
\item{[\cantat]} S. Cantat, {\sl Sur les groupes de transformations birationnelles des surfaces}, Ann. Math. {\bf 174} (2011) 299-340.
\item{[\arnold]} V. I. Arnold, {\sl Dynamics of complexity of intersections}, Bol. Soc. Bras. Mat. {\bf 21} (1990) 1-10.
\item{[\veselov]} A.P. Veselov, {\sl Growth and integrability in the dynamics of mappings}, Commun. Math. Phys. {\bf 145} (1992) 181-193.
\item{[\takenawa]} T. Takenawa, {\sl Algebraic entropy and the space of initial values for discrete dynamical systems}, J. Phys. A: Math. Gen. {\bf 34} (2001) 10533-10545.
\item{[\bedkim]} E. Bedford and K. Kim, {\sl On the degree growth of birational mappings in higher dimension}, J. Geom. Anal. {\bf 14} (2004) 567-596.
\item{[\belletal]} J.P. Bell, J. Diller, M. Jonsson and H. Kriger, {\sl Birational maps with transcendental dynamical degree}, proc. Londono Math. Soc. {\bf 3} (2024) 128:e12573.
\item{[\alonsoo]}  J. Alonso, Yu.B. Suris and Kangning Wei, {\sl Dynamical degrees of birational maps from indices of polynomials with respect to blow-ups I. General theory and 2D examples} (2023) arXiv:2303.15864 [math.DS].
\item{[\alonsot]}  J. Alonso, Yu.B. Suris and Kangning Wei, {\sl Dynamical degrees of birational maps from indices of polynomials with respect to blow-ups II. 3D examples} (2023) arXiv:2307.09939 [math.DS].
\item{[\vialleto]} C-M. Viallet, {\sl On the degree growth of iterated birational maps}, arXiv:1909.13259 [math.AG].
\item{[\viallett]} C-M. Viallet, {\sl An exercise in experimental mathematics: calculation of the algebraic entropy of a map}, Open Comm. Nonl. Math. Phys. ]ocnmp[ Special Issue {\bf 1} (2024) pp 1-11.
\item{[\hardrod]} R.G. Halburd, {\sl Elementary calculations of degree growth and entropy for discrete equations}, Proc. R. Soc. A. {\bf 473} (2017) 20160831.
\item{[\express]} A. Ramani, B. Grammaticos, R. Willox and T. Mase, {\sl Calculating algebraic entropies: an express method}, J. Phys. A {\bf 50} (2017) 185203.
\item{[\maserod]} T. Mase, R.Willox, A. Ramani and B. Grammaticos, {\sl Singularity confinement as an integrability
criterion} J. Phys. A {\bf 52} (2019) 205201.
\item{[\nonconfrod]} A. Ramani , B. Grammaticos , R. Willox , T. Mase and J. Satsuma, {\sl Calculating the algebraic entropy of mappings with unconfined singularities}, J. Int. Sys. {\bf 3} (2018) xyy006.
\item{[\higher]} A. Ramani, B. Grammaticos, A.S. Carstea and R. Willox, {\sl Obtaining the growth of higher order mapping through the study of singularities}, J. Phys. A: Math. Theor. {\bf 58} (2025) 115201.
\item{[\premierpapier]} R. Willox, T. Mase, A. Ramani and B. Grammaticos, {\sl Singularities and growth of higher order discrete equations}, Op. Comm. Nonlin. Math. Phys, Special Issue {\bf 2} (2024) 46-64. 
\item{[\diophantine]} R. G. Halburd, {\sl Diophantine integrability}, J. Phys. A: Math. Gen. {\bf 38} (2005) L1-L7.
\item{[\silvo]} J.H. Silverman, {\sl Dynamical degrees, arithmetic entropy, and canonical heights for dominant rational self-maps of projective space}, Erg. Theor. Dyn. Syst. {\bf 34} (2012) 647-678.
\item{[\kawasilv]} S. Kawaguchi and J.H. Silverman, {\sl On the dynamical and arithmetic degrees of rational self-maps of algebraic varieties},  J. Reine Angew. Math. {\bf 713} (2016) p. 21-48.
\item{[\silvt]} J.H. Silverman, {\sl Arithmetic and Dynamical Degrees on Abelian Varieties}, J. Th\'eorie des Nombres de Bordeaux {\bf 29} (2017), 151-167.
\item{[\matsu]} Y. Matsuzawa, K. Sano and T. Shibata, {\sl Arithmetic degrees for dynamical systems over function fields of characteristic zero}, Math. Z. {\bf 290} (2018) 1063-1083. 
\item{[\eulerbook]} B. Grammaticos, A. Ramani, R. Willox and T. Mase, {\sl Detecting discrete integrability: the singularity approach}, in {\it Nonlinear Systems and Their Remarkable Mathematical Structures: Volume I}, N. Euler (Ed.) (CRC Press, Boca Raton FL, 2018) (pp 44--73).
\item{[\asymm]} M.D. Kruskal, K.M. Tamizhmani, B. Grammaticos and A. Ramani, {\sl Asymmetric discrete Painlev\'e equations}, Reg. Chaot. Dyn. {\bf 5} (2000) 273-280.
\item{[\fast]} B. Grammaticos, A. Ramani, A.S. Carstea and R. Willox, {\sl A fast algorithmic way to calculate the degree growth of birational mappings}, Mathematics {\bf 13} (2025) 737.
\item{[\anticonf]} T. Mase, R. Willox, B. Grammaticos and A. Ramani, {\sl Integrable mappings and the notion of anticonfinement}, J. Phys. A: Math. Theor. {\bf 51} (2018) 265201.
\item{[\toki]} T. Tokihiro, D. Takahashi, J. Matsukidaira and J. Satsuma, {\sl  From Soliton Equations to Integrable Cellular Automata through a Limiting Procedure}, Phys. Rev. Lett. {\bf 76} (1996) 3247-3250.
\item{[\limproc]}  A. Ramani, B. Grammaticos and Y. Ohta, {\sl Discrete integrable systems from continuous Painlev\'e equations through limiting procedures}, Nonlinearity {\bf 13} (2000) 1073-1085.
\item{[\takenawahiv]} T. Takenawa, {\sl Discrete Dynamical Systems Associated with Root Systems of Indefinite Type},
Commun. Math. Phys. {\bf 224} (2001) 657-681.
\item{[\hiv]} J. Hietarinta and C-M. Viallet, {\sl Singularity Confinement and Chaos in Discrete Systems}, Phys. Rev. Lett. {\bf 81} (1998) 325-328.
\item{[\takcarst]} A.S. Carstea and T. Takenawa, {\sl Space of initial conditions and geometry of two 4-dimensional discrete Painlev\'e equations}, J. Phys. A: Math. Theor. {\bf 52} (2019)  275201.
\item{[\alextakcarst]} A. Stokes, T. Takenawa and A.S. Carstea, {\sl On the geometry of a 4-dimensional extension of a $q$-Painlev\'e I equation with symmetry type $A_1^{(1)}$}, J. Phys. A: Math. Theor. {\bf 58} (2025) 405201.
\item{[\late]} B. Grammaticos, A. Ramani, R. Willox, T. Mase and J. Satsuma, {\sl Singularity confinement and full-deautonomisation: a discrete integrability criterion},  Physica D {\bf 313} (2015) 11-25.
\item{[\joshiviallet]} N. Joshi and C-M. Viallet, {\sl Rational Maps with Invariant Surfaces}, J. Int. Syst. {\bf 3} (2018)  xyy017.

\end